\documentclass[preprint,11pt,3p]{elsarticle}
\usepackage{amssymb,amsmath,graphicx,bbm,subfigure}
\begin{document}
\title{Geometric interpretation of the geometric discord}
\author{Yao Yao}
\author{Hong-Wei Li}
\author{Zhen-Qiang Yin\corref{cor1}}
\ead{yinzheqi@mail.ustc.edu.cn}
\author{Zheng-Fu Han\corref{cor2}}
\ead{zfhan@ustc.edu.cn}
\cortext[cor1]{Corresponding author}
\cortext[cor2]{Principal corresponding author}
\address{Key Laboratory of Quantum Information,University of Science and
Technology of China,Hefei 230026,China}
\date{\today}

\begin{abstract}
We investigate the level surfaces of geometric measure of quantum discord, and provide a pictorial
interpretation of geometric discord for Bell-diagonal states. We have observed its nonanalytic behavior
under decoherence employing this approach and interestingly found if we expect geometric discord to remain constant
under phase-flip channel for a finite period, the initial state must be separable. Besides, this geometric
understanding can be applied to verify the hierarchical relationships between geometric discord and the original one.
The present work makes us conjecture that the incompatibility of these two definitions may
originate from the discrepancy of the geometric structures of them.
\end{abstract}


\maketitle


\section{Introduction}
Entanglement, as a crucial resource for quantum information theory, plays a vital role in quantum
information processes and its fundamental nonclassical aspect is well recognized \cite{horodecki}.
However, it has recently been demonstrated that entanglement is not the only aspect of quantum
correlations both theoretically and experimentally \cite{without enanglement}. Consequently, the
conventional entanglement-separability framework seems to be inappropriate in the sense of
characterizing and quantifying quantum correlations. Therefore, how to qualify and quantify quantumness of correlations
remains an open question for us.

Aiming to capture the total nonclassical correlations, Ollivier and Zurek introduced a measure called quantum
discord \cite{discord}, which have received a great deal of attention lately \cite{bell state,x state,general state,mid,unify,
geometric discord,geometry of bell,sudden transition,gmqd,nonzero bloch,decoherence1,decoherence2,non-markovian,ferraro}.
They observed two classically identical expressions for the mutual information differ in a quantum case, and realized this
difference can be utilized to depict the quantumness of correlations. Consider a composite bipartite sysytem $\rho^{AB}$, and let
$\rho^{A(B)}=Tr_{B(A)}(\rho^{AB})$ denote the reduced density operator of the part A(B). The quantum mutual information
is defined as
\begin{eqnarray}
\mathcal{I}(\rho^{AB})&:=S(\rho^{A})+S(\rho^{B})-S(\rho^{AB})\\
                      &=S(\rho^{A})-S(\rho^{AB}|\rho^{B}),
\end{eqnarray}
where $S(\rho)=-Tr(\rho\log_2\rho)$ is the von Neumann entropy. Moreover, it has been shown that quantum mutual
information captures the total correlation of a bipartite quantum system. In order to quantify quantum discord,
Ollivier and Zurek took a alternative way to generalize the classical mutual information by use of a
measurement-based conditional density operator. If $\{\Pi^{B}_{k}\}$ denotes a set of one-dimensional von Neumann
measurement performed on subsystem B, then the resulting state conditioned on the measurement outcome labeled by k
is
\begin{equation}
\rho_{k}=\frac{1}{p_{k}}(I^{A}\otimes\Pi^{B}_{k})\rho(I^{A}\otimes\Pi^{B}_{k}),
\end{equation}
where probability $p_{k}=Tr[(I^{A}\otimes\Pi^{B}_{k})\rho]$, and $I^{A}$ is the identity operator for part A.
Note that it was already proven in \cite{povm} for two-qubit states the projective measurement is always the optimal
choice for the conditional entropy, and hence the application of generic positive operator-valued measure (POVM)
is not necessary. The quantum conditional entropy according to this measurement is defined as
\begin{equation}
S(\rho|\{\Pi^{B}_{k}\}):=\sum_{k}p_{k}S(\rho_{k}),
\end{equation}
and the generalized quantum mutual information with respect to this measurement yields
\begin{equation}
\mathcal{I}(\rho|\{\Pi^{B}_{k}\}):=S(\rho^{A})-S(\rho|\{\Pi^{B}_{k}\}),
\end{equation}
By optimizing over all possible von Neumann measurement $\{\Pi^{B}_{k}\}$, the quantity
\begin{equation}
\mathcal{J}(\rho):=\sup_{\{\Pi^{B}_{k}\}}\mathcal{I}(\rho|\{\Pi^{B}_{k}\}),
\end{equation}
can be regarded as a measure of classical information. The discrepancy between the original quantum mutual
information $\mathcal{I}$ and the measurement-induced quantum mutual information $\mathcal{J}$ is defined
as the so called quantum discord
\begin{equation}
\mathcal{D}(\rho):=\mathcal{I}(\rho)-\mathcal{J}(\rho).
\end{equation}
which captures the total quantum correlation. Due to the complicated optimization procedure for calculating
the classical correlation, the evaluation of quantum discord is a tough task from a computational point of
view. Up to now, the analytical expression for quantum discord is only available for Bell-diagonal states \cite{bell state} and
a certain class of X-structured states \cite{x state}. The most recent work \cite{general state} reveals that a closed expression
for the discord of arbitrary states of two qubits cannot be obtained. Significantly, this difficulty in
computing quantum discord motivated the proposals of alterative definitions of quantum correlations in turn.

As early as in \cite{mid}, Luo pointed out that a natural consideration leads us to a measure of quantum correlations
as follows
\begin{equation}
\mathcal{Q}_{D}(\rho):=\inf_{\Pi}D(\rho,\Pi(\rho)),
\label{distance}
\end{equation}
where the infimum is taken with respect to all complete local projective measurements $\Pi$, and the reasonable
candidates of $D(\cdot,\cdot)$ are the trace distance, the Hilbert-Schmidt distance, the Bures distance, or even
the relative entropy (which is a pseudodistance). Based on the concept of relative entropy, Modi et al. presented
a unified view of quantum and classical correlations, and this scenario puts all correlations on an equal footing
\cite{unify}. In addition, Dakic et al. introduced the following geometric measure of quantum discord using the
Hilbert-Schmidt norm \cite{geometric discord}
\begin{equation}
\mathcal{D}_{G}(\rho):=\min_{\chi\in\Omega}\|\rho-\chi\|^2,
\label{HS-norm}
\end{equation}
where $\Omega$ denotes the set of zero-discord states and $\|\rho-\chi\|^2=Tr(\rho-\chi)^2$ is the square of
Hilbert-Schmidt norm of Hermitian operators. It is worth emphasizing that, based on a simplified definition of
the geometric discord in Ref. \cite{gmqd} by Luo and Fu, the definition Eq. (\ref{HS-norm}) is also a particular
case of Eq. (\ref{distance}). One can write an arbitrary two-qubit state in the Bloch decomposition
\begin{eqnarray}
\label{two-qubit}
\rho=\frac{1}{4}(I^{A}\otimes I^{B}+\sum^3_{i=1}x_i\sigma_i\otimes I^{B}+\sum^3_{i=1}y_i I^{A}\otimes\sigma_i
+\sum^3_{i,j=1}T_{ij}\sigma_i\otimes\sigma_j),
\end{eqnarray}
where $x_i=Tr\rho(\sigma_i\otimes I^{B})$, $y_i=Tr\rho(I^{A}\otimes\sigma_i)$ are components of the local Bloch vectors, $T_{ij}=Tr\rho(\sigma_i\otimes\sigma_j)$ are components of the correlation tensor, and $\sigma_i,i\in\{1,2,3\}$ are the three
Pauli matrices. The geometric measure of quantum discord of Eq. (\ref{two-qubit}) can be evaluated as \cite{geometric discord}
\begin{equation}
\mathcal{D}_{G}(\rho)=\frac{1}{4}(\|\vec{x}\|^2+\|T\|-k_{max}).
\end{equation}
where $\vec{x}:=(x_1,x_2,x_3)^T$ is a column vector, $T:=(t_{ij})$ is a matrix, and $k_{max}$ is the largest eigenvalue of
the matrix $K=\vec{x}\vec{x}^T+TT^T$.

Interestingly, M. D. Lang and C. M. Caves \cite{geometry of bell} considered the level serfaces of quantum discord for Bell-diagonal states, and provided a pictorial approach which presented a complete interpretation of the strcture of quantum discord and its
dynamic behavior under decoherence. As indicated in \cite{geometry of bell}, the phenomenon (sudden transition between classical and quantum decoherence) investigated in \cite{sudden transition} can be easily seen from the surfaces of constant discord. Inspired by this scenario,
it is natural to generalize the use of the geometrical method to the geometric discord. As one might expect, we gain an keen
insight into the dynamics of the geometric discord.

The remainder of the paper is arranged as follows. In Sec. II, we give a brief review on the quantum discord and geometric discord
with respect to Bell-diagonal states and then investigate the level surfaces of the geometric discord. In Sec. III, we evaluate
the dynamic behavior of the geometric discord under decoherence channel based on this pictorial approach, and also verify the hierarchical relationships between the geometric discord and the original one. Finally, Sec. IV is devoted to the discussion and conclusion.
\section{Level surfaces of geometric discord}
\begin{figure}[htbp]
\begin{center}
\includegraphics[width=.40\textwidth]{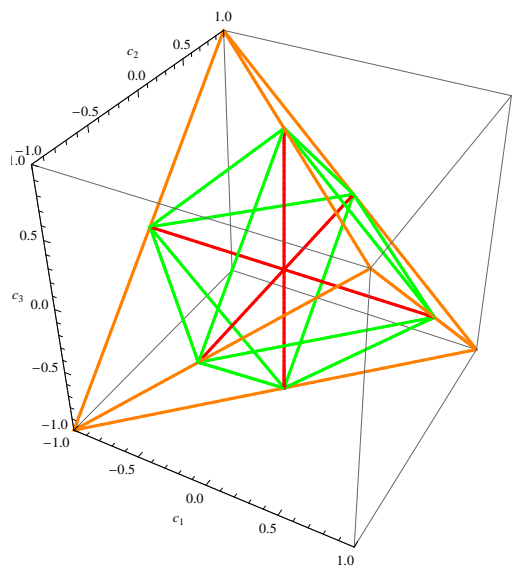} {}
\end{center}
\caption{(Color online) Geometrical representation of Bell-diagonal states: the orange-line-contoured tetrahedron
represents valid Bell-diagonal states, the green-line-contoured octahedron denotes separable states, which is actually
bounded by $|c_1|+|c_2|+|c_3|\leq1$ and the zero-discord states are labeled by the red three lines (Cartesian axes).
This geometrical picture clearly displays that almost all states have nonzero discord \cite{ferraro}.
}\label{Bell}
\end{figure}

We begin with the two-qubit Bell-diagonal states, which have density operators of the form \cite{geometry of bell}
\begin{eqnarray}
\rho&=\frac{1}{4}(I^A\otimes I^B+\sum^{3}_{i=1}c_i\sigma^A_i\otimes\sigma^B_i)
=\sum_{a,b=0,1}\lambda_{a,b}|\beta_{a,b}\rangle\langle\beta_{a,b}|,
\end{eqnarray}
where the eigenstates are four Bell states $|\beta_{a,b}\rangle\equiv(|0,b\rangle+(-1)^a|1,1\oplus b\rangle)/\sqrt{2}$
with eigenvalues
\begin{equation}
\lambda_{a,b}=\frac{1}{4}(1+(-1)^{a}c_1-(-1)^{a+b}c_2+(-1)^{b}c_3).
\end{equation}
Note that each state $\rho$ is associated with a 3-tuple $(c_1,c_2,c_3)$, and this state is physical (positive operator)
if $\lambda_{ab}\geq0$, that is
\begin{eqnarray}
\label{positivity}
1-c_1+c_2+c_3\geq0,\,1+c_1-c_2+c_3\geq0,\nonumber\\
1+c_1+c_2-c_3\geq0,\,1-c_1-c_2-c_3\geq0,
\end{eqnarray}
Clearly, the above conditions show that the vector $\vec{c}=(c_1,c_2,c_3)$ belongs to the tetrahedron $\mathcal{T}$ with vertices
$(-1,-1,-1)$, $(-1,1,1)$, $(1,-1,1)$, and $(1,1,-1)$ (Fig. \ref{Bell}). A Bell-diagonal state is separable if and only if its
partial transpose is positive. In matrix form, we can rewrite
\begin{equation}
\label{Bell-diagonal}
\rho=\frac{1}{4}
\left(\begin{array}{cccc}
1+c_3 & 0 & 0 & c_1-c_2 \\
0 & 1-c_3 & c_1+c_2 & 0 \\
0 & c_1+c_2& 1-c_3 & 0 \\
c_1-c_2 & 0 & 0 & 1+c_3
\end{array}\right),
\end{equation}
\begin{equation}
\rho^{T_B}=\frac{1}{4}
\left(\begin{array}{cccc}
1+c_3 & 0 & 0 & c_1+c_2 \\
0 & 1-c_3 & c_1-c_2 & 0 \\
0 & c_1-c_2& 1-c_3 & 0 \\
c_1+c_2 & 0 & 0 & 1+c_3
\end{array}\right).
\end{equation}
It can easily be seen that the partial transpose just flips the sign of $c_2$, thus the physical region of $\rho^{T_B}$
is the reflection of $\mathcal{T}$ through the plane $c_2=0$. Finally, the resulting region of separable Bell-diagonal
states is the intersection of the two tetrahedra, which belongs to the octahedron $\mathcal{L}$ with vertices
$(\pm1,0,0)$, $(0,\pm1,0)$, $(0,0,\pm1)$ (Fig. \ref{Bell}) \cite{inseparability}.

To investigate the difference of level surfaces between quantum discord and geometric discord, let us first recall
the analytic formulas of them. For original measure of quantum discord, the quantum mutual information is given by \cite{bell state}
\begin{equation}
\mathcal{I}(\rho)=2-S(\rho^{AB})=2+\sum_{a,b}\lambda_{ab}\log_2\lambda_{ab},
\end{equation}
the classical correlation is given by
\begin{eqnarray}
\mathcal{J}(\rho)=1-H_2(\frac{1+c}{2})
=\frac{1+c}{2}\log_2(1+c)+\frac{1-c}{2}\log_2(1-c),
\end{eqnarray}
where $H_2(x)=-x\log_{2}x-(1-x)\log_{2}(1-x)$, and $c=\max\{|c_1|,|c_2|,|c_3|\}$. Therefore, the quantum discord is the
difference of $\mathcal{I}$ and $\mathcal{J}$
\begin{eqnarray}
\mathcal{D}(\rho)=&\frac{1}{4}[(1-c_1-c_2-c_3)\log_2(1-c_1-c_2-c_3)\nonumber\\
&+(1-c_1+c_2+c_3)\log_2(1-c_1+c_2+c_3)\nonumber\\
&+(1+c_1-c_2+c_3)\log_2(1+c_1-c_2+c_3)\nonumber\\
&+(1+c_1+c_2-c_3)\log_2(1+c_1+c_2-c_3)]\nonumber\\
&-\frac{1+c}{2}\log_2(1+c)-\frac{1-c}{2}\log_2(1-c).
\end{eqnarray}

On the other hand, the geometric measure of quantum discord can be obtained explicitly as \cite{geometric discord}
\begin{equation}
\mathcal{D}_G(\rho)=\frac{1}{4}(c_1^2+c_2^2+c_3^2-\max\{c_1^2,c_2^2,c_3^2\}),
\end{equation}
One can easily find function $\mathcal{D}_G(\rho)=\mathcal{D}_G(c_1,c_2,c_3)$ possesses symmetry properties
\begin{eqnarray}
\mathcal{D}_G(c_1,c_2,c_3)&=\mathcal{D}_G(\pm c_i,\pm c_j,\pm c_k)
\,for\,i\neq j\neq k,
\end{eqnarray}

\begin{figure}[htbp]
\centering
\subfigure[]{
\includegraphics[width=0.21\textwidth]{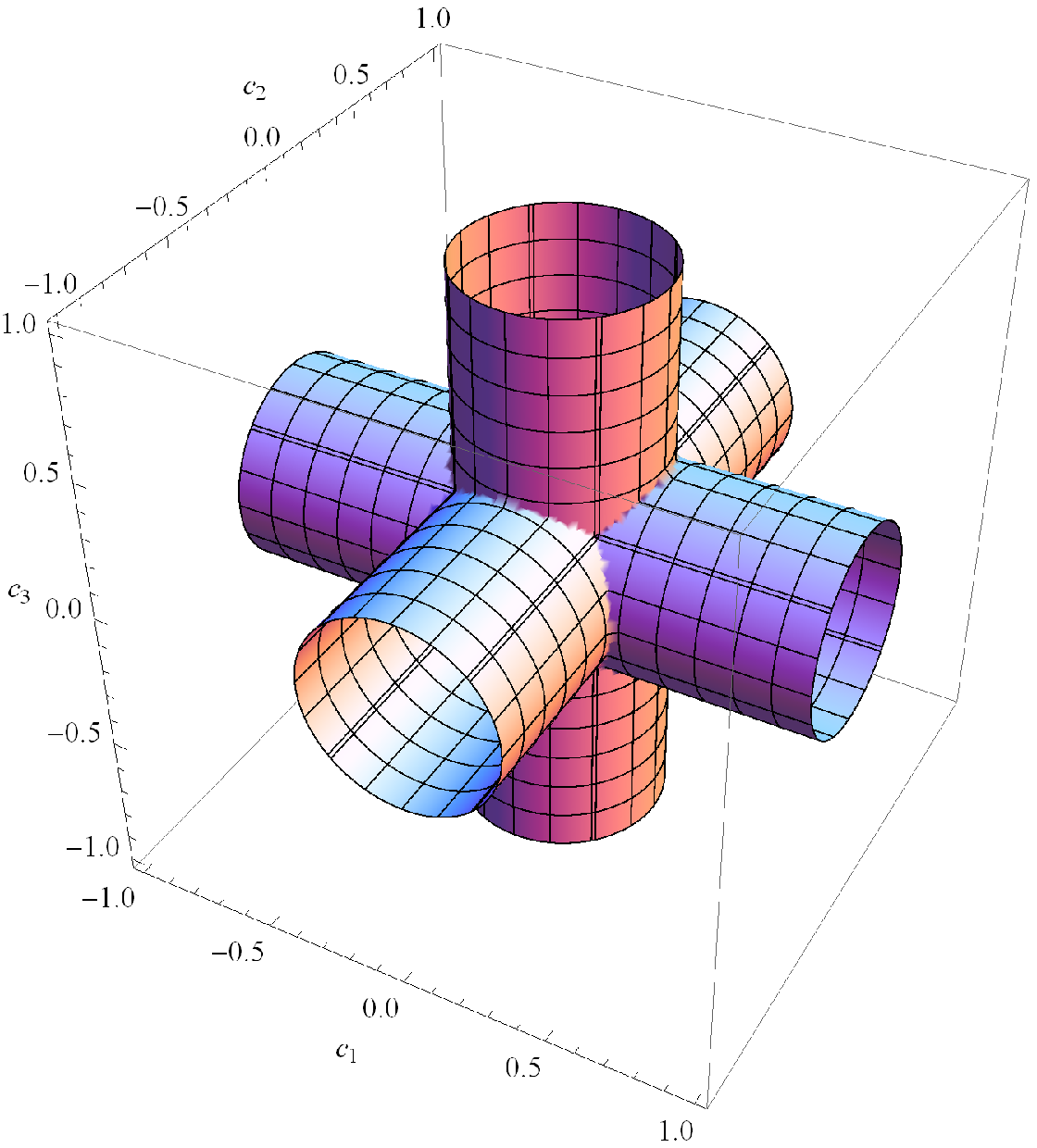}
\includegraphics[width=0.21\textwidth]{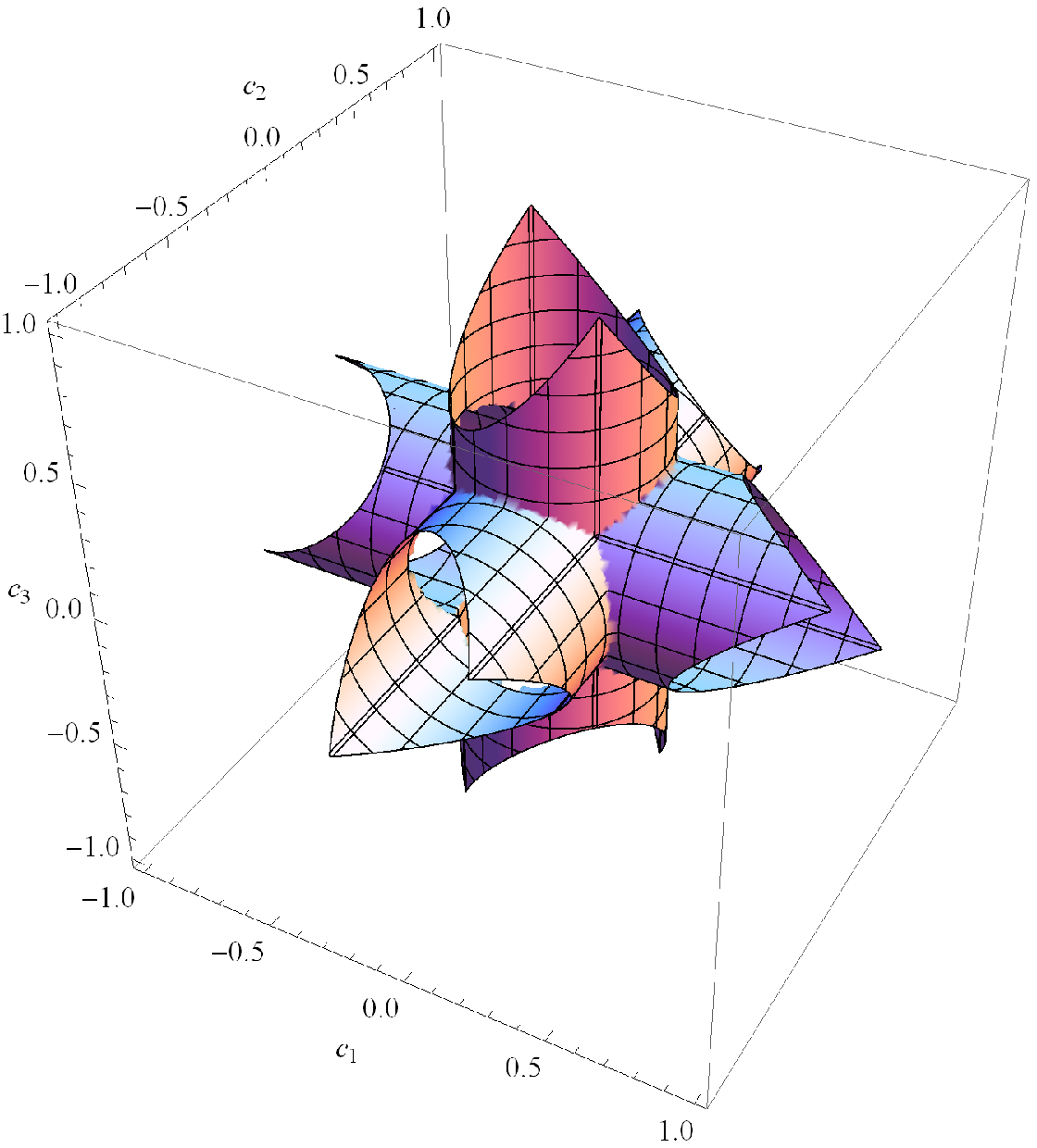}\label{a}}
\subfigure[]{
\includegraphics[width=0.21\textwidth]{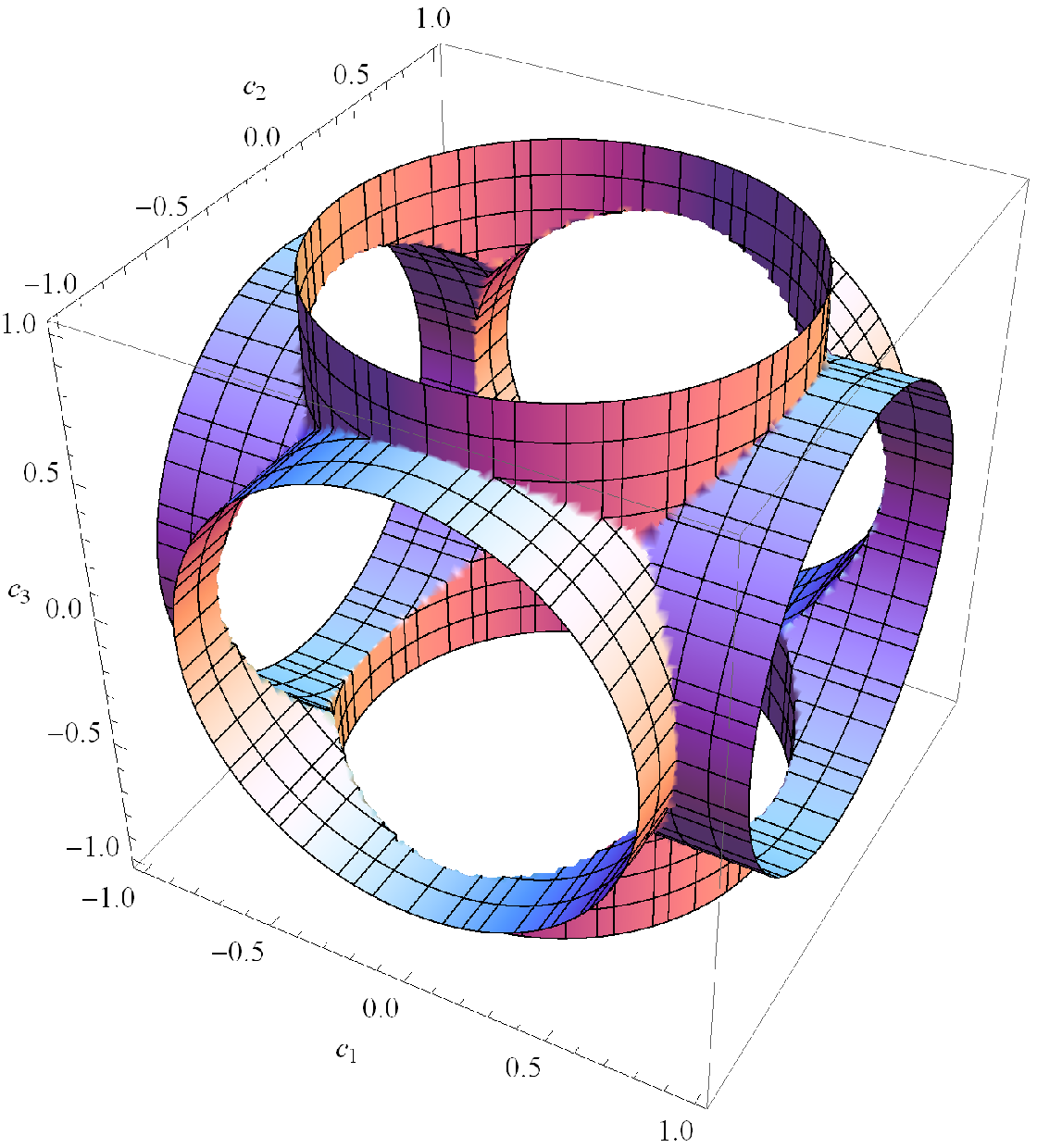}
\includegraphics[width=0.21\textwidth]{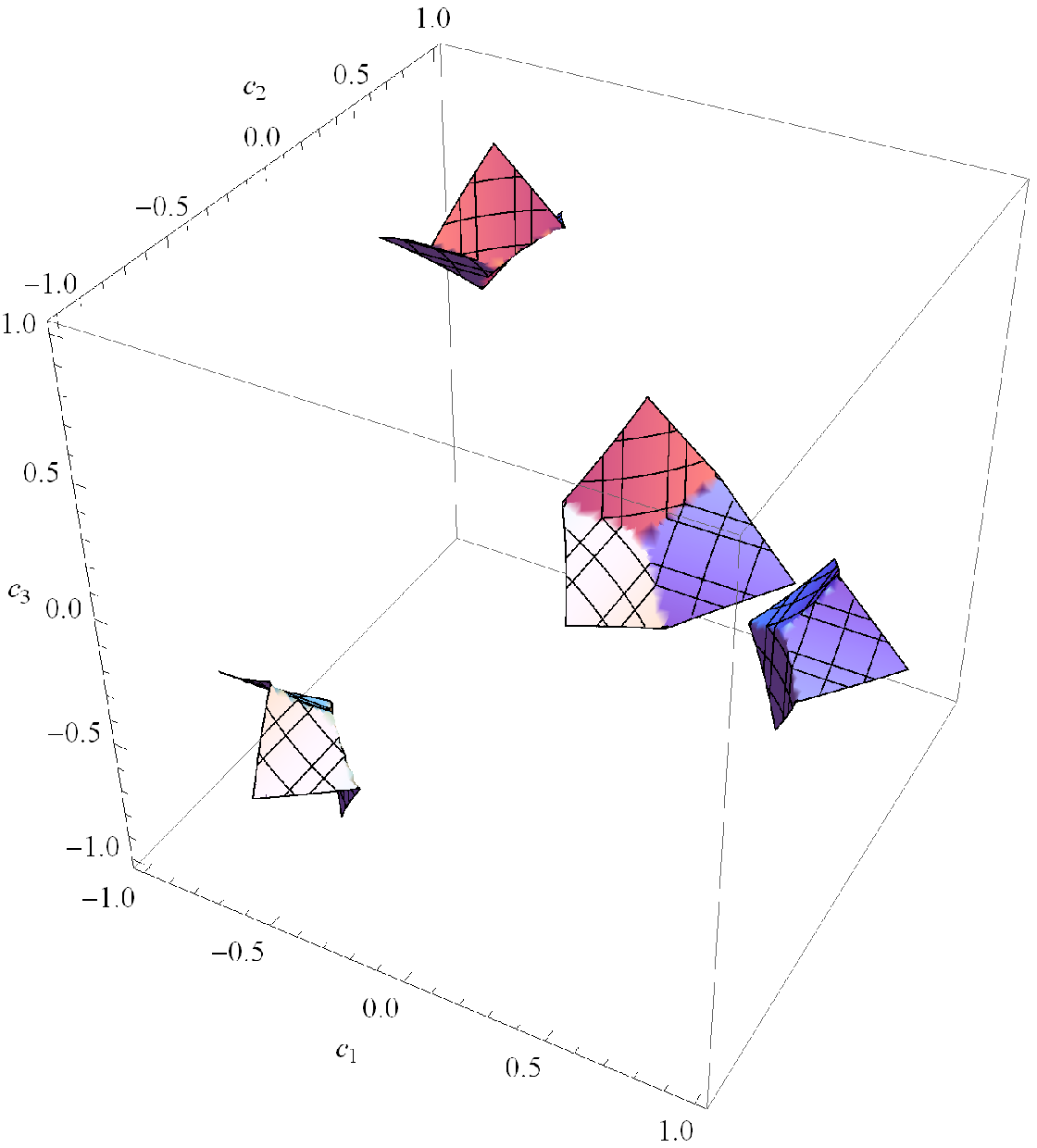}\label{b}}
\subfigure[]{
\includegraphics[width=0.21\textwidth]{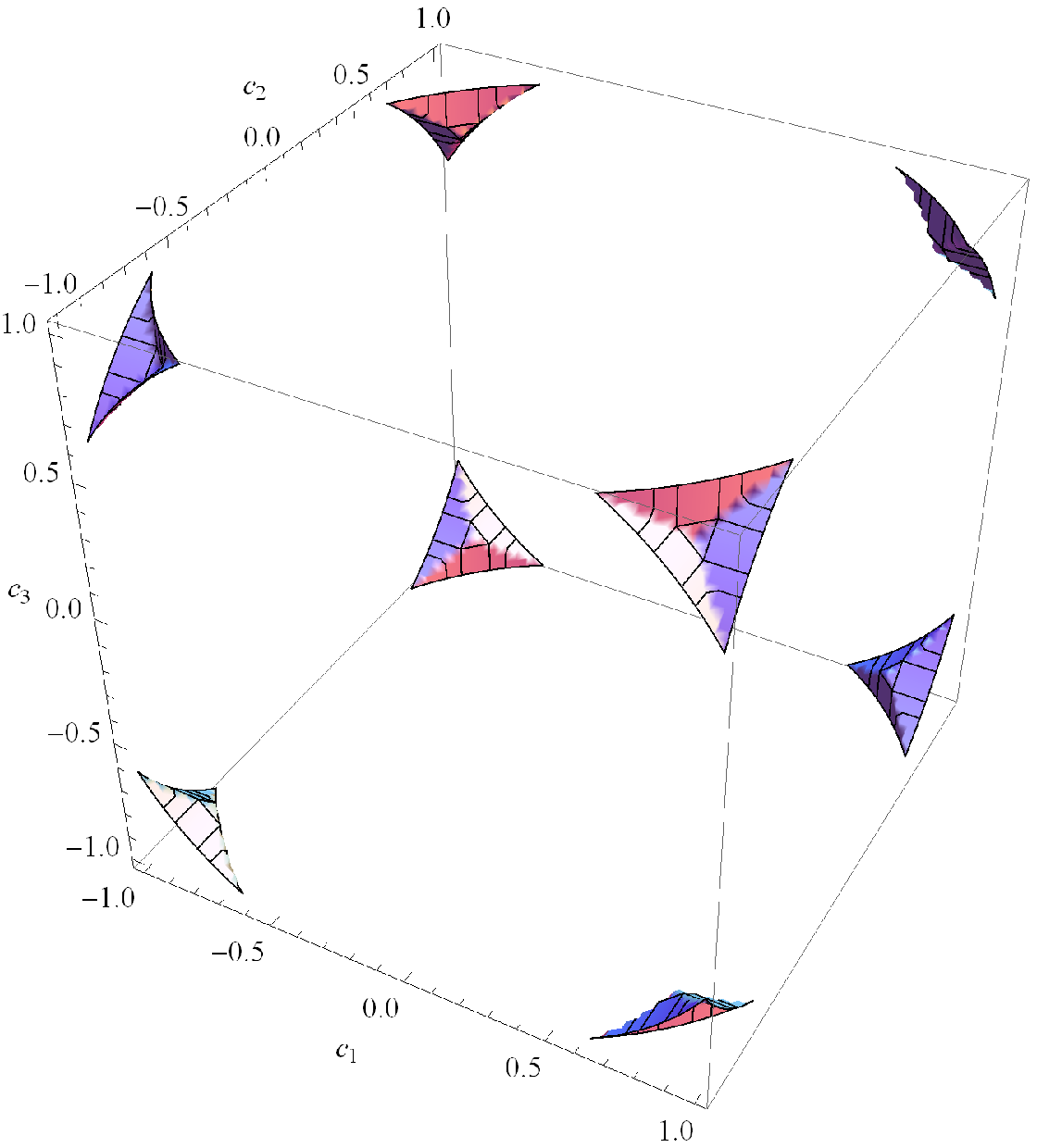}
\includegraphics[width=0.21\textwidth]{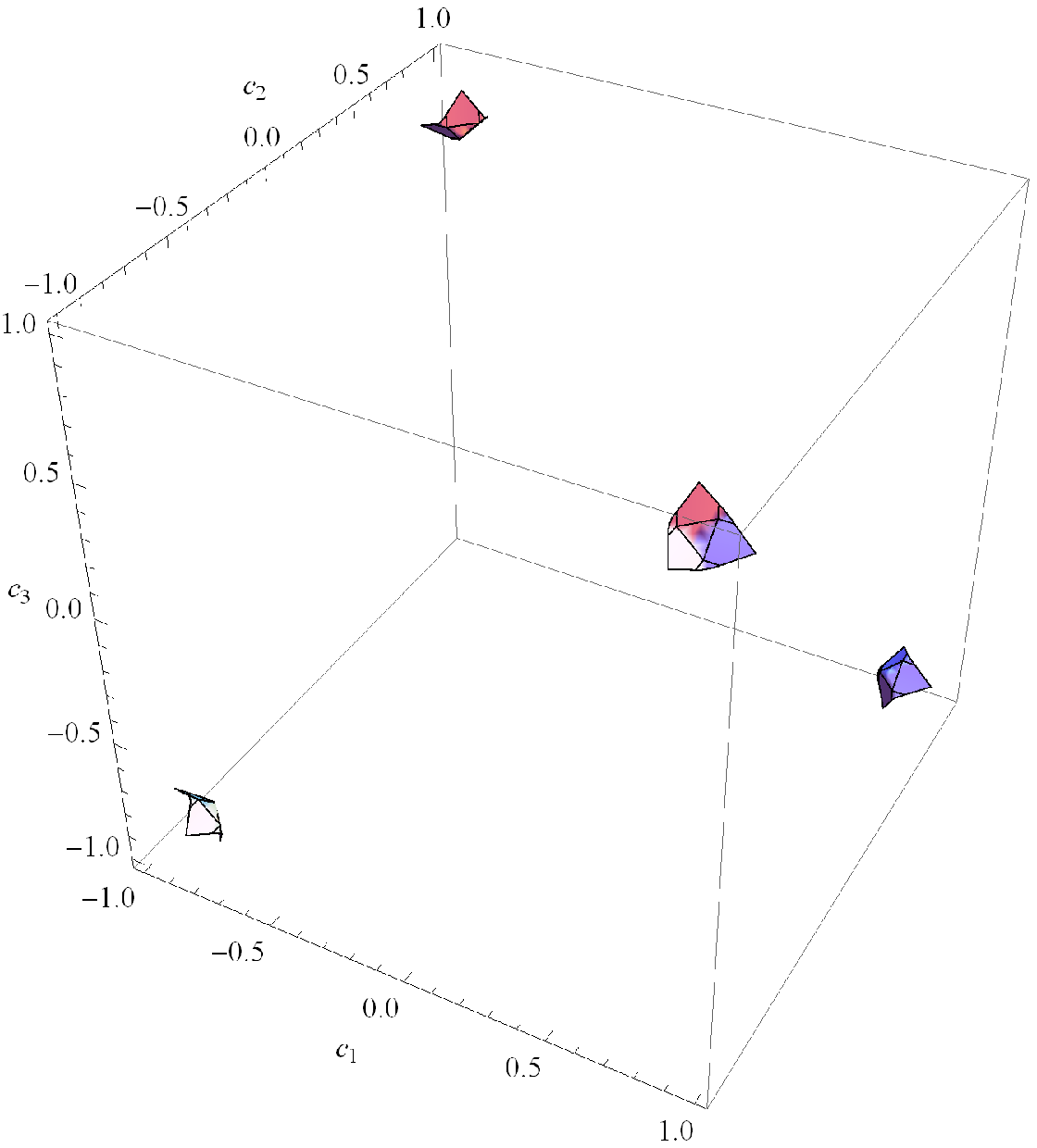}\label{c}}
\caption{(Color online) Level surfaces of constant geometric discord: (a) $\mathcal{D}_G=0.03$,
(b) $\mathcal{D}_G=0.15$, (c) $\mathcal{D}_G=0.35$. In each subfigure, left picture shows the original
contour of geometric discord, while right graph illustrates the level surface in consideration of state
tetrahedron $\mathcal{T}$.
}\label{surface1}
\end{figure}

In particular, $\mathcal{D}_G(c_1,c_2,c_3)=\mathcal{D}_G(-c_1,-c_2,-c_3)$, which means the level surfaces
(without consideration of $\mathcal{T}$) are symmetric with respect to planes $c_1=0$, $c_2=0$, and $c_3=0$,
as showed in Fig. \ref{surface1}. Apparently, $\mathcal{D}(c_1,c_2,c_3)=\mathcal{D}(c_i,c_j,c_k)$ still holds
for quantum discord, but it is usually not the case when we flip the sign of arbitrary variables.

Compared with level surfaces of quantum discord depicted in \cite{geometry of bell}, the level surfaces of geometric discord
are composed of three identical intersecting "cylinders" instead of irregular "tubes" (see Fig. \ref{surface1}).
The cylinders are running along the three Cartesian axes, and their ends are cut off by the valid state
tetrahedron $\mathcal{T}$. The cylinders shrink towards the Cartesian axes as geometric discord becomes smaller;
meanwhile, if discord increases the structure of cylinders is cut into four identical pieces reaching out
towards the vertices of $\mathcal{T}$, which stand for the four Bell states. The phenomenon is similar to
geometrical picture of quantum discord.

Recently, we notice that the geometrical depiction of quantum discord is also investigated for a family of
two-qubit states with parallel nonzero Bloch vectors \cite{nonzero bloch}
\begin{eqnarray}
\rho=\frac{1}{4}(I\otimes I+r\sigma_3\otimes I+I\otimes s\sigma_3+\sum^{3}_{i=1}c_i\sigma^A_i\otimes\sigma^B_i),
\end{eqnarray}
where the parameters $r$ and $s$ are real constants. One can also give the matrix form
\begin{equation}
\label{deform}
\rho=\frac{1}{4}
\left(\begin{array}{cccc}
1+r+s+c_3 & 0 & 0 & c_1-c_2 \\
0 & 1+r-s-c_3 & c_1+c_2 & 0 \\
0 & c_1+c_2& 1-r+s-c_3 & 0 \\
c_1-c_2 & 0 & 0 & 1-r-s+c_3
\end{array}\right),
\end{equation}

It is usually complicated to evaluate the quantum discord for such a class of states due to optimization
procedure \cite{nonzero bloch}. However, we can obtain the geometric discord through simple calculation
\begin{equation}
\mathcal{D}_G(\rho)=\frac{1}{4}(c_1^2+c_2^2+c_3^2+r^2-\max\{c_1^2,c_2^2,c_3^2+r^2\}),
\end{equation}
It is worth noting that when the Bloch vectors are nonzero (that is, $r,s\neq0$), the geometric objects
$\mathcal{T}$ and $\mathcal{L}$ will be deformed \cite{deformation}. The eigenvalues of $\rho$ in Eq. (\ref{deform}) are
given by
\begin{eqnarray}
\mu_{\pm}=\frac{1}{4}[(1-c_3)\pm\sqrt{(r-s)^2+(c_1+c_2)^2}],\\
\nu_{\pm}=\frac{1}{4}[(1+c_3)\pm\sqrt{(r+s)^2+(c_1-c_2)^2}].
\end{eqnarray}
Hence the deformation of $\mathcal{T}$ can be demonstrated concerning the positivity property of $\rho$.
In fact, the constraint condition of the deformation turns into \cite{deformation}
\begin{equation}
\min(\mu_-,\nu_-)=0.
\end{equation}

\begin{figure}[!htbp]
\centering
\subfigure[]{
\includegraphics[width=0.2\textwidth]{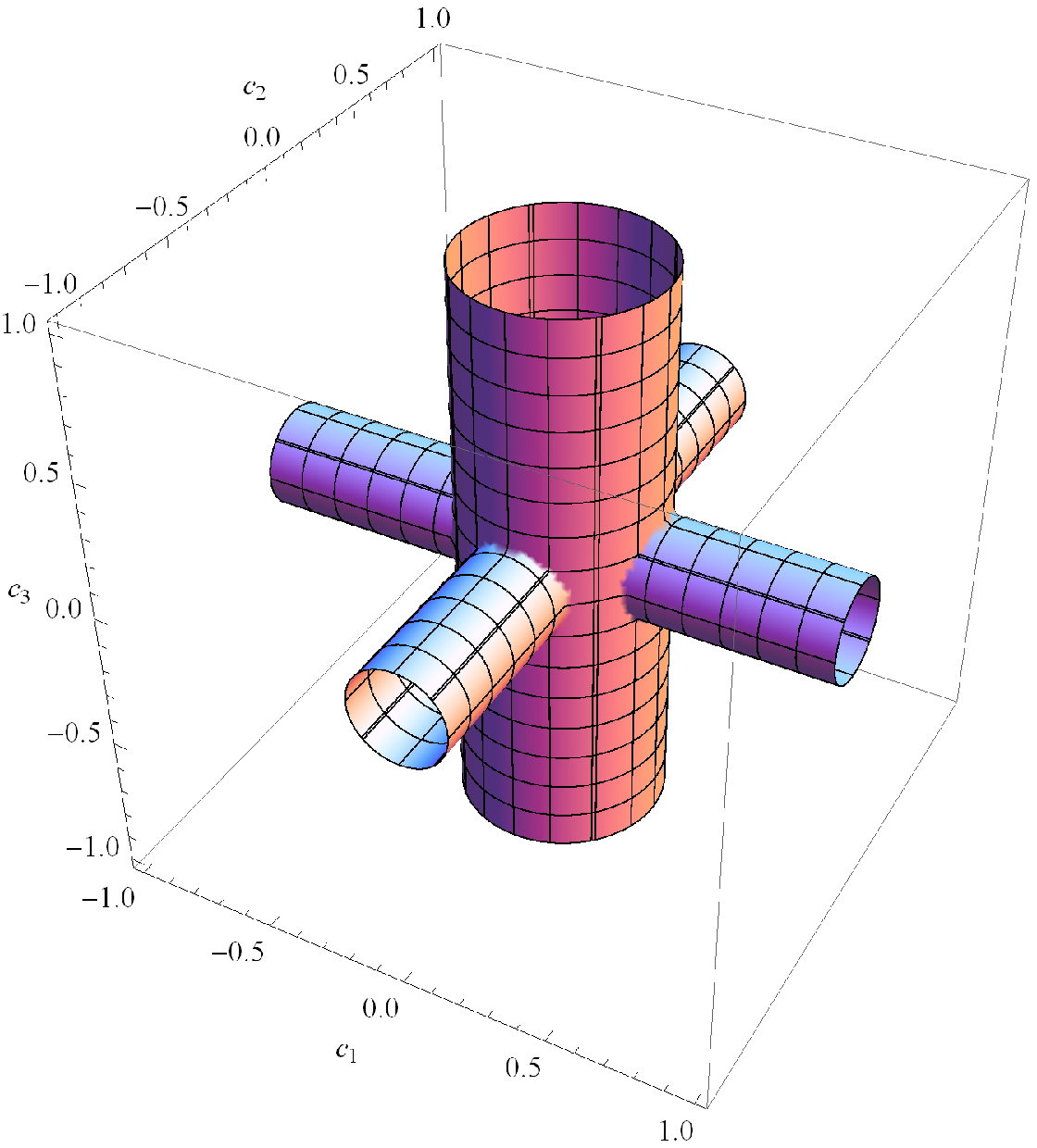}
\includegraphics[width=0.2\textwidth]{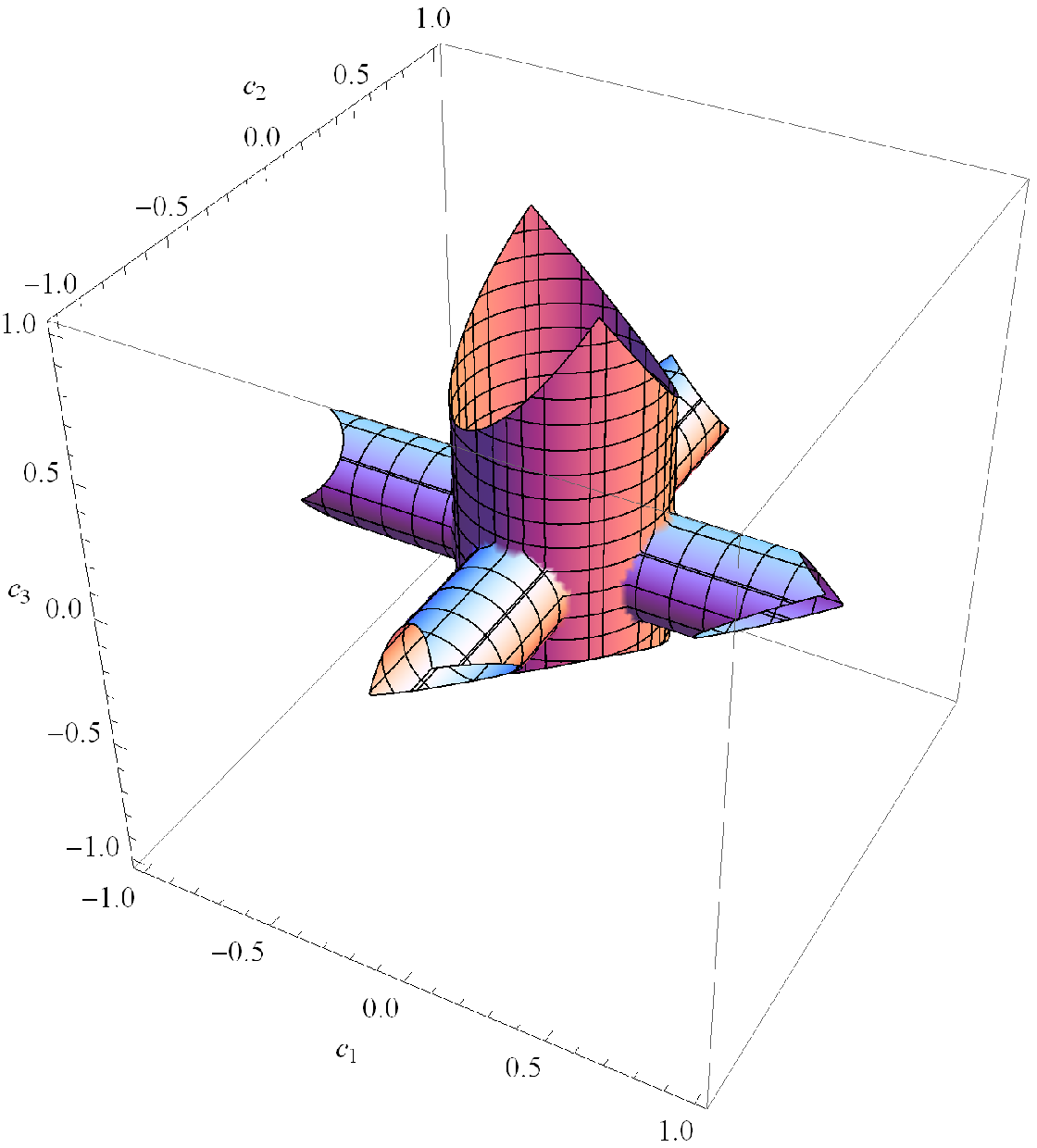}\label{a}}
\subfigure[]{
\includegraphics[width=0.2\textwidth]{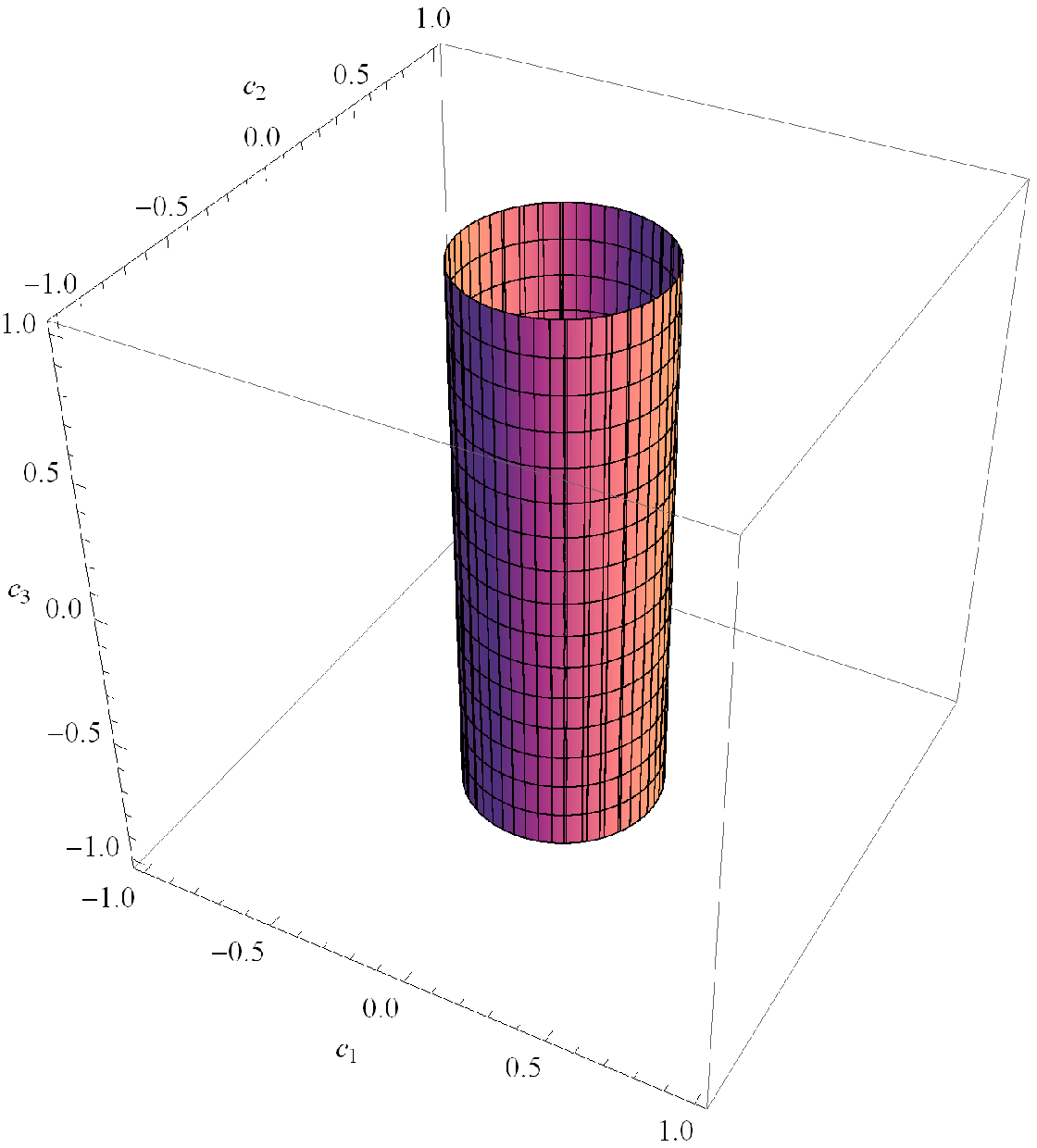}
\includegraphics[width=0.2\textwidth]{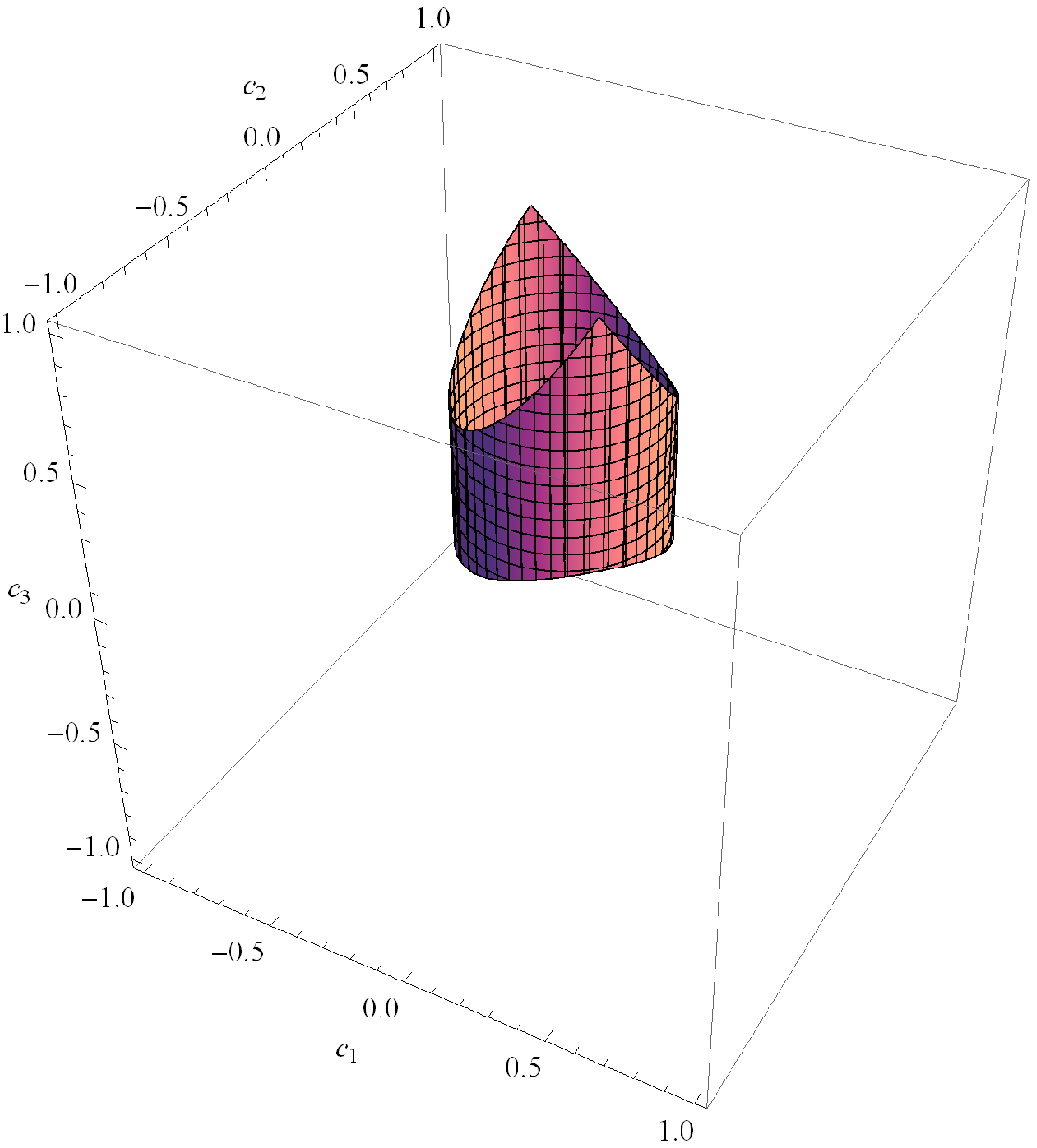}\label{b}}
\subfigure[]{
\includegraphics[width=0.2\textwidth]{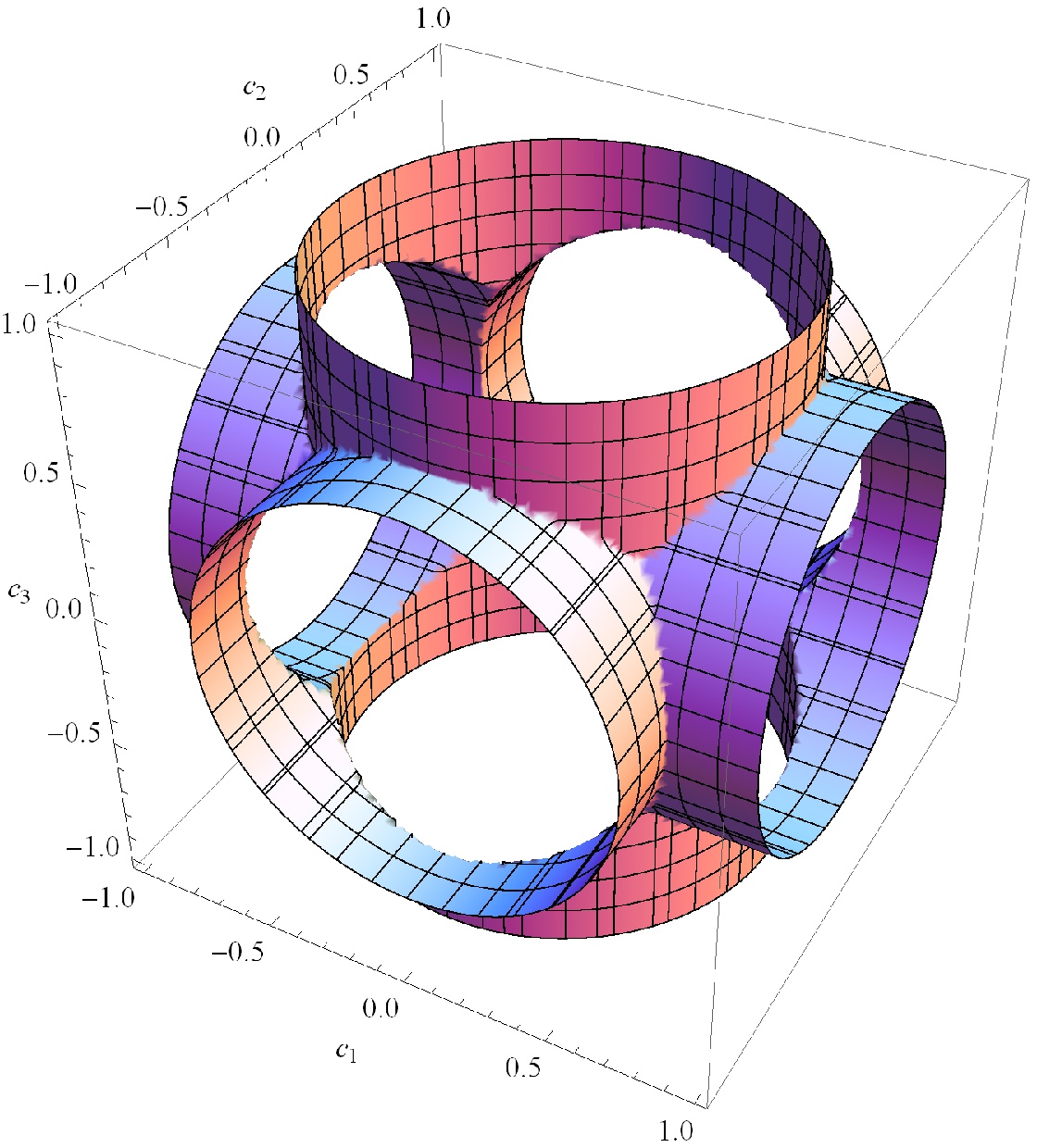}
\includegraphics[width=0.2\textwidth]{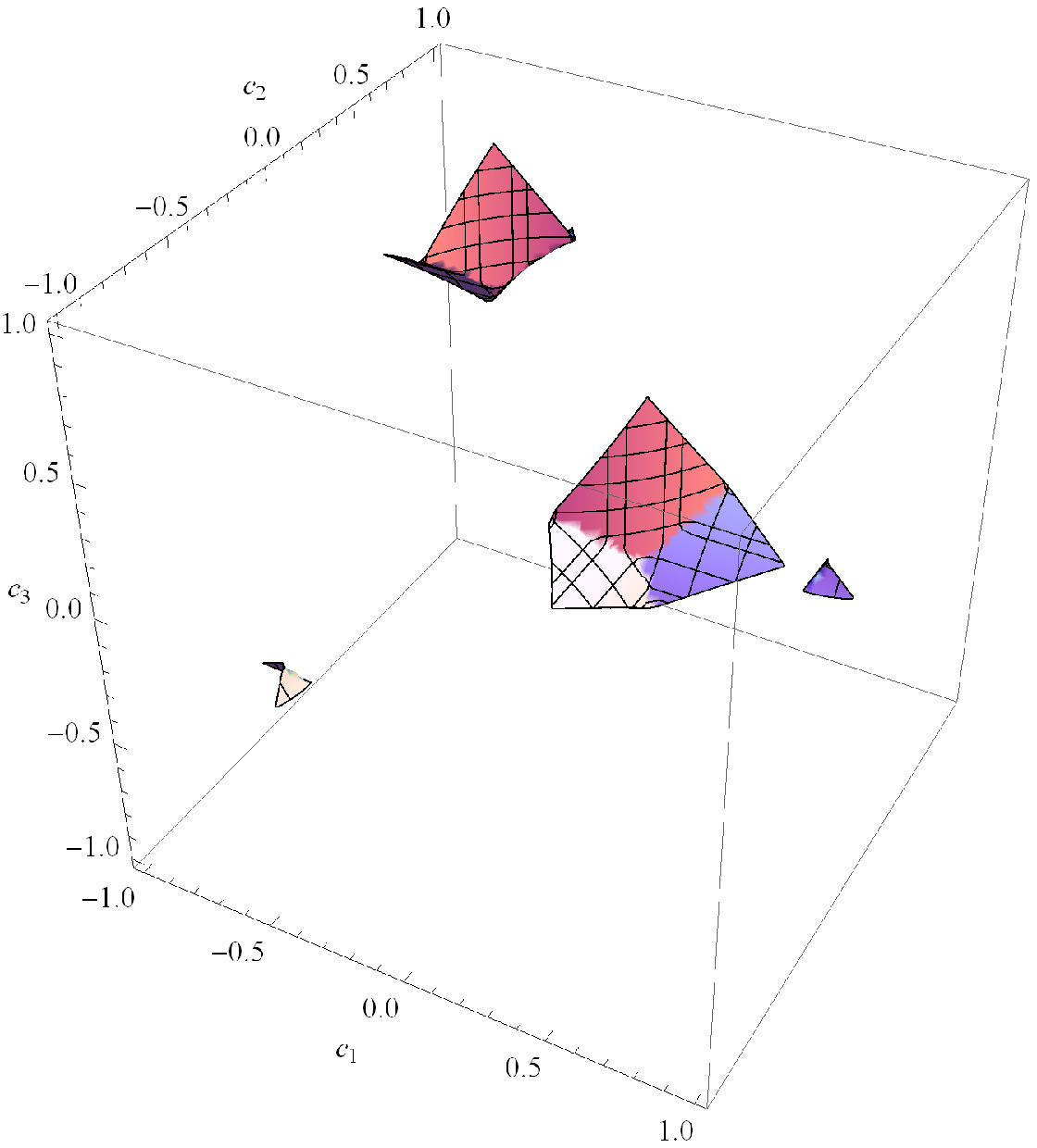}\label{c}}
\subfigure[]{
\includegraphics[width=0.2\textwidth]{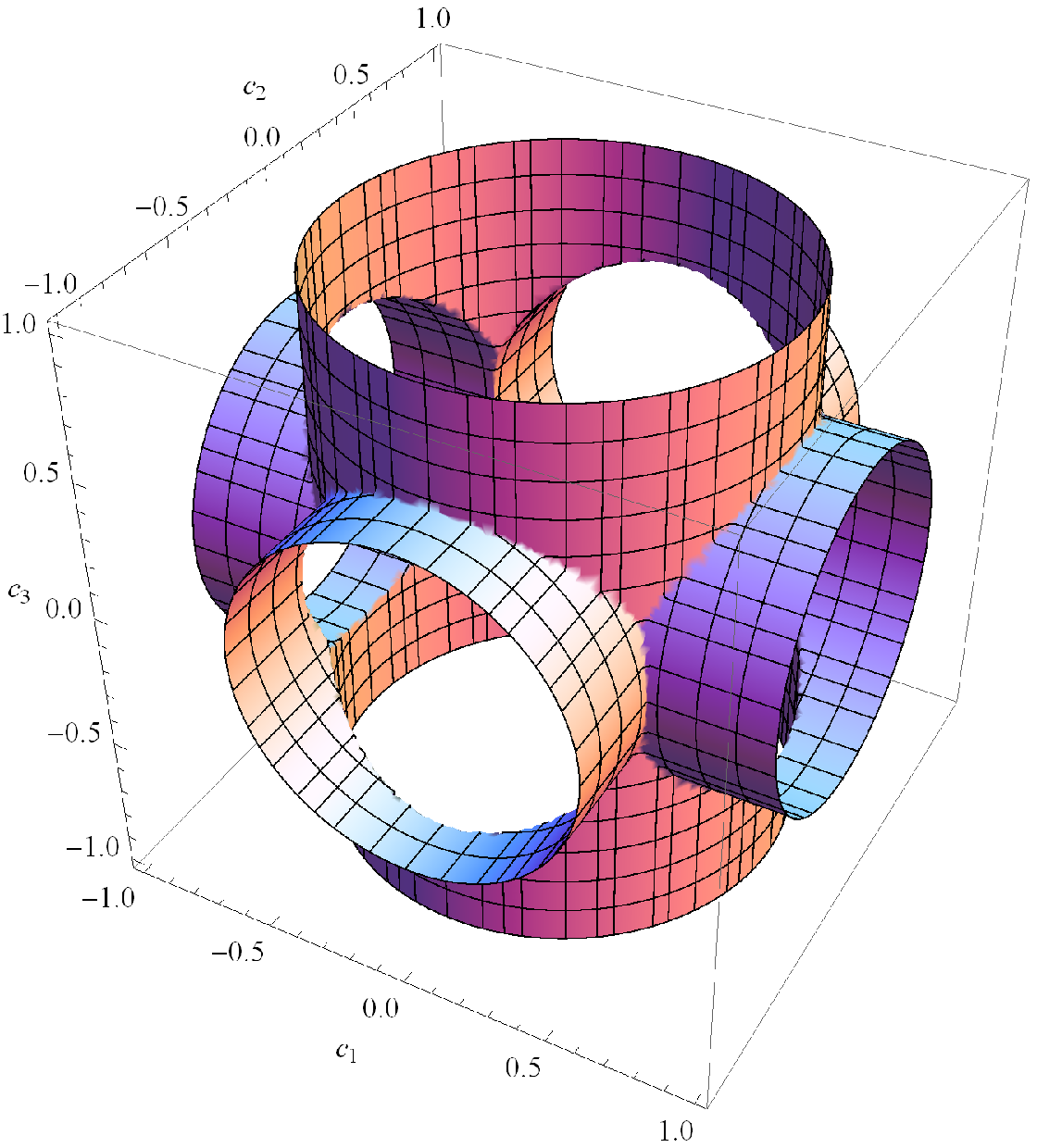}
\includegraphics[width=0.2\textwidth]{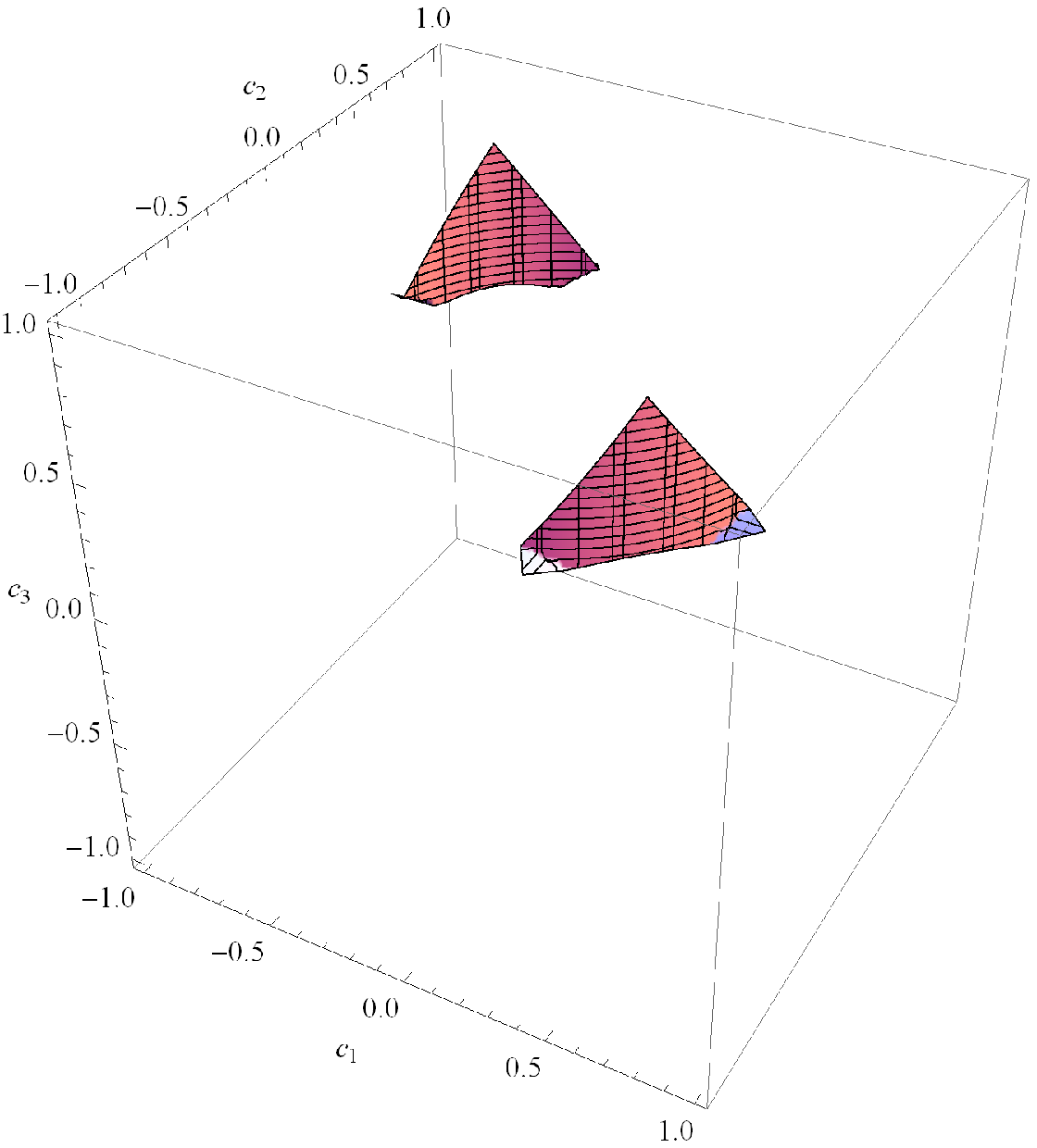}\label{d}}
\caption{(Color online) Surfaces of constant geometric discord of state $\rho$ defined in Eq. (\ref{deform}):
(a) $r=s=0.3$, $\mathcal{D}_G=0.03$, (b) $r=s=0.5$, $\mathcal{D}_G=0.03$,
(c) $r=s=0.3$, $\mathcal{D}_G=0.15$, (d) $r=s=0.5$, $\mathcal{D}_G=0.15$. In each subfigure, left picture shows the original
contour of geometric discord, while right graph illustrates the level surface considering the geometrical deformation of
tetrahedron $\mathcal{T}$.
}\label{surface2}
\end{figure}

In Fig. \ref{surface2}, we plot level surfaces of of geometric discord of state $\rho$ defined in Eq. (\ref{deform}).
From these figures, one can see that the level surfaces in this case are quite different from the ones represented
in \cite{nonzero bloch}: though the surfaces here also shrinks with the change of $r$ and the shrinking rate becomes larger with
increasing $r$, the horizontal cylinder-like "tubes" are not closed when geometric discord gets smaller
[see Fig. \ref{surface2}(a)], and moreover, for large $r$ the original contour never moves above the plane $c_3=0$
[Fig. \ref{surface2}(c)(d)] because $\mathcal{D}_G(c_1,c_2,c_3)=\mathcal{D}_G(-c_1,-c_2,-c_3)$ still holds in this
situation. Finally, we should emphasize that level surfaces of geometric discord shrink along with deformation of
$\mathcal{T}$. In Fig. \ref{surface2}, one can observe that if we consider the deformation, the surfaces is biased
toward the $(v_1,v_2)$ region ($v_1=(1,-1,1)$, $v_2=(-1,1,1)$), which is compatible with the results in \cite{deformation}.

\section{Applications of level surfaces}
\subsection{Geometric discord under decoherence}
Recently, several authors \cite{decoherence1,decoherence2} have investigated the dissipative dynamics of two-qubit quantum discord under local Markovian environments. Following the same method, that is, the Kraus operator approach \cite{kraus}, it is convenient to
obtain analytic formulas to describe the evolution of the geometric discord under decoherence. Given an initial
state $\rho$ for two qubits A and B, its evolution can be modeled in the Kraus representation
\begin{equation}
\varepsilon(\rho)=\sum_{i,j}E_{i,j}\rho(0)E_{i,j}^{\dag},
\end{equation}
where $E_{i,j}=E^A_i\otimes E^B_j$ are Kraus operators, satisfying $E_{i,j}^{\dag}E_{i,j}=I$ if the quantum operation
is trace-preserving, and the operators $E_{i(j)}$ denotes the one-qubit decoherence effects. Here, we give the Kraus
operators for some typical kinds of decoherence channels \cite{nielsen,channels}, and calculate the geometric discord for Bell-diagonal
states under each channel (only the final results are shown for simplicity).

\textit{Amplitude damping channel:} Kraus operators $E_0=diag\{1,\sqrt{1-p}\}$, $E_1=\sqrt{p}(\sigma_1+i\sigma_2)/2$.
\begin{eqnarray}
\varepsilon(\rho)=\frac{1}{4}[I+p(\sigma_3\otimes I+I\otimes\sigma_3)+qc_1\sigma_1\otimes\sigma_1
+qc_2\sigma_2\otimes\sigma_2+(p^2+c_3q^2)\sigma_3\otimes\sigma_3],
\end{eqnarray}
\begin{eqnarray}
\mathcal{D}_G=\frac{1}{4}[q^2(c_1^2+c_2^2)+(p^2+c_3q^2)^2+p^2
-\max\{(qc_1)^2,(qc_2)^2,(p^2+c_3q^2)^2+p^2\}],
\end{eqnarray}
where $q=1-p$, and the same below.

\textit{Phase damping channel:} Kraus operators $E_0=diag\{1,\sqrt{1-p}\}$, $E_1=diag\{0,\sqrt{p}\}$.
\begin{eqnarray}
\varepsilon(\rho)=\frac{1}{4}[I+qc_1\sigma_1\otimes\sigma_1+qc_2\sigma_2\otimes\sigma_2+c_3\sigma_3\otimes\sigma_3],
\end{eqnarray}
\begin{eqnarray}
\mathcal{D}_G=\frac{1}{4}[q^2(c_1^2+c_2^2)+c_3^2-\max\{(qc_1)^2,(qc_2)^2,c_3^2\}],
\end{eqnarray}

\textit{Depolarizing channel:} Kraus operators $E_0=\sqrt{1-3p/4}I\}$, $E_1=\sqrt{p/4}\sigma_1$,
$E_2=\sqrt{p/4}\sigma_2$, $E_3=\sqrt{p/4}\sigma_3$.
\begin{eqnarray}
\varepsilon(\rho)=\frac{1}{4}[I+q^2c_1\sigma_1\otimes\sigma_1+q^2c_2\sigma_2\otimes\sigma_2+q^2c_3\sigma_3\otimes\sigma_3],
\end{eqnarray}
\begin{eqnarray}
\mathcal{D}_G=\frac{1}{4}[q^4(c_1^2+c_2^2+c_3^2)-\max\{(q^2c_1)^2,(q^2c_2)^2,(q^2c_3)^2\}],
\end{eqnarray}

\textit{Bit flip, phase flip, and bit-phase flip channel:} $E_0=\sqrt{1-p/2}I$, $E_1^i=\sqrt{p/2}\sigma_i$, where
$i=1$ gives us the bit flip, $i=2$ the bit-phase flip, and $i=3$ the phase flip.
\begin{eqnarray}
\varepsilon(\rho)=\frac{1}{4}[I+q^2c_j\sigma_j\otimes\sigma_j+q^2c_k\sigma_k\otimes\sigma_k+c_i\sigma_i\otimes\sigma_i],
\end{eqnarray}
\begin{eqnarray}
\mathcal{D}_G=\frac{1}{4}[q^4(c_j^2+c_k^2)+c_i^2-\max\{(q^2c_j)^2,(q^2c_k)^2,c_i^2\}].
\end{eqnarray}
where $i\neq j\neq k$, and $i,j,k=1,2,3$.

Note that the parametrized time p is responsible for a wide range of physical phenomena \cite{nielsen}. For example, for the
dephasing channel (the phase damping and phase flip channels are actually the same quantum operation \cite{nielsen}, which can
also be seen from the calculations above), $p=1-\exp(-\Gamma t)$ with $\Gamma$ the decay rate \cite{open system}.

Mazzola et al. \cite{sudden transition} recently observed that with certain initial conditions, quantum discord remains constant
for a finite time interval even under decoherence. Thus one is natually led to ask this question: whether
such a situation will happen to the geometric discord? To pursue the previous works,
we focus on the dynamics of geometric discord under independent phase flip
(or phase damping) channels for the two qubits. We already have the expression for $\mathcal{D}_G$
\begin{eqnarray}
\mathcal{D}_G(p)=\frac{1}{4}[c_1^2(p)+c_2^2(p)+c_3^2-\max\{c_1^2(p),c_2^2(p),c_3^2\}],
\end{eqnarray}
where $c_1(p)=(1-p)^2c_1(0)$, $c_2(p)=(1-p)^2c_2(0)$, and $c_3$ remains unchanged ($0\leq p<1$).
Without loss of generality, we assume the initial conditions $|c_1(0)|\geq|c_2(0)|,|c_3(0)|$. when the
time goes on, the geometric discord can be presented as
\begin{eqnarray}
\mathcal{D}_G(p)=\left\{\begin{array}{cc}
\frac{1}{4}(c_2^2(p)+c_3^2), &  \, p\leq 1-\sqrt{|c_3|/|c_1|}\\
\frac{1}{4}(c_1^2(p)+c_2^2(p)), & \, p>1-\sqrt{|c_3|/|c_1|}
\end{array}\right.
\end{eqnarray}
Therefore, if we do not expect the geometric discord to be spoiled for a time period, then $c_2^2(p)$
must keep constant, that is, $c_2(0)=0$. In this case, $\mathcal{D}_G=c_3^2/4$.

\begin{figure}[htbp]
\begin{center}
\includegraphics[width=.40\textwidth]{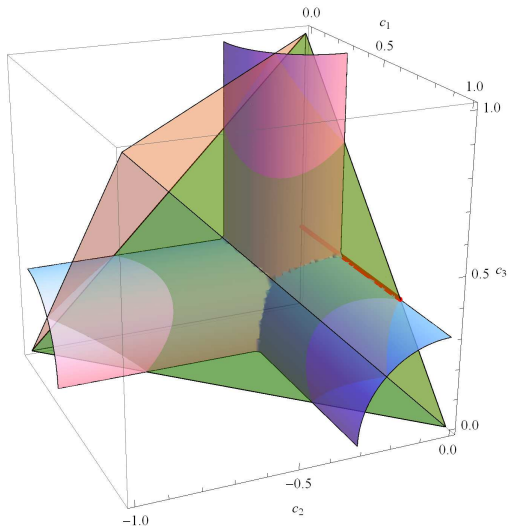} {}
\end{center}
\caption{(Color online) The trajectory (red solid line) of a Bell-diagonal state under local phase flip channels
(see context for more details). Only the $(+,-,+)$ region is show, and the orange state tetrahedron $\mathcal{T}$
and the green separable octahedron $\mathcal{L}$ are also plotted. A level surface of constant geometric discord
is displayed as a background with the initial conditions $c_1(0)=0.6$, $c_2(0)=0$, $c_3(0)=0.3$.
}\label{surface3}
\end{figure}

In fact, we can gain a more detailed insight into this problem by use of the geometric interpretation
of the geometric discord. In Fig. \ref{surface3}, we demonstrate the trajectory traced out by the
decohering-state taking the level surface as the background. The straight line $\{c_1(p)=(1-p)^2c_1(0),c_2(p)=0,c_3(p)=c_3\}$
lies on the plane $c_2=0$ and runs along the generating line of "cylinder", until it comes across the vertical "cylinder"
at $c_1(p)=c_3$, which implies a "sudden change" point of geometric discord. Subsequently, the geometric discord
decreases monotonically to zero till the trajectory arrives at the $c_3$ axis, when the state becomes completely classical.

It is truly remarkable that, illustrated by the pictorial approach, the initial state must be separable! To identify
this intuitive knowledge, we adopt Wootters's "concurrence" \cite{concurrence} to calculate the entanglement. For Bell-diagonal
state described in Eq. (\ref{Bell-diagonal}) (X-structured state), one can easily obtain
\begin{eqnarray}
C(\rho)=2\max\{0,\Lambda_1,\Lambda_2\},
\end{eqnarray}
where $\Lambda_1=|\frac{c_1-c_2}{4}|-|\frac{1-c_3}{4}|$, $\Lambda_2=|\frac{c_1+c_2}{4}|-|\frac{1+c_3}{4}|$.
Since the geometric discord can remain undestroyed under phase flip channels if and only if $c_2(0)=0$
with initial conditions $|c_1|\geq|c_2|,|c_3|$, the positivity Eq. (\ref{positivity}) reduces to
\begin{eqnarray}
|c_1+c_3|\leq1,\,|c_1-c_3|\leq1,
\end{eqnarray}
So we obtain $|c_1|+|c_3|=\max\{|c_1+c_3|,|c_1-c_3|\}\leq1$. Employing the triangle property $|1\pm c_3|\geq1-|c_3|$,
one can easily verify
\begin{eqnarray}
\Lambda_1,\Lambda_2\leq\frac{1}{4}(|c_1|+|c_3|-1)\leq0,
\end{eqnarray}
That is to say, $C(\rho)=0$, which means the initial state must be separable.

\begin{figure}[!htbp]
\begin{center}
\subfigure[]{
\label{a}
\includegraphics[width=0.3\textwidth ]{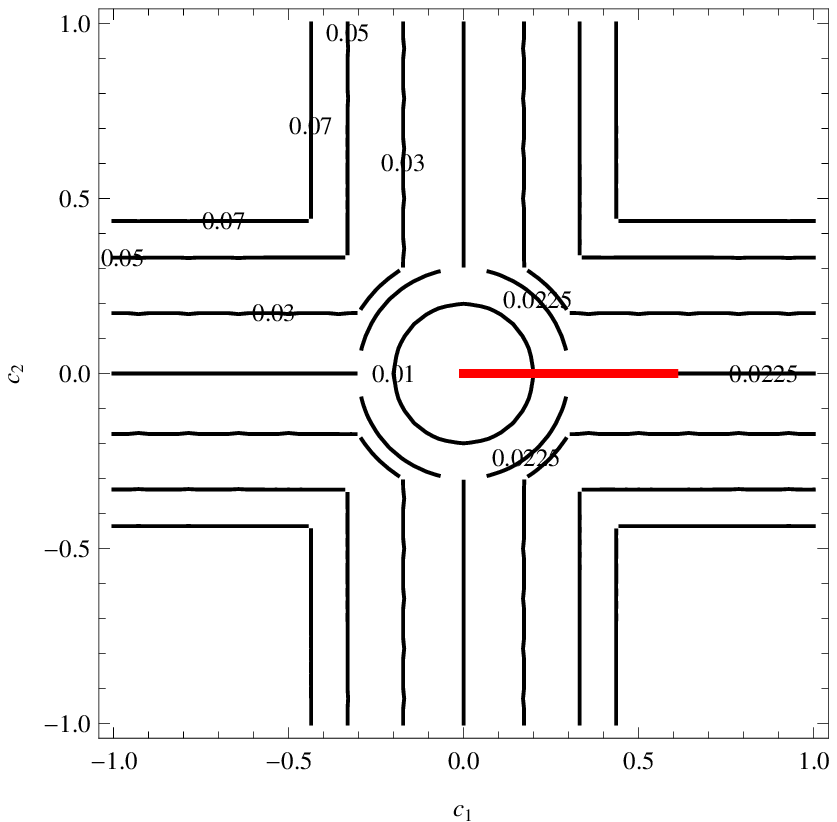}}
\subfigure[]{
\label{b}
\includegraphics[width=0.3\textwidth ]{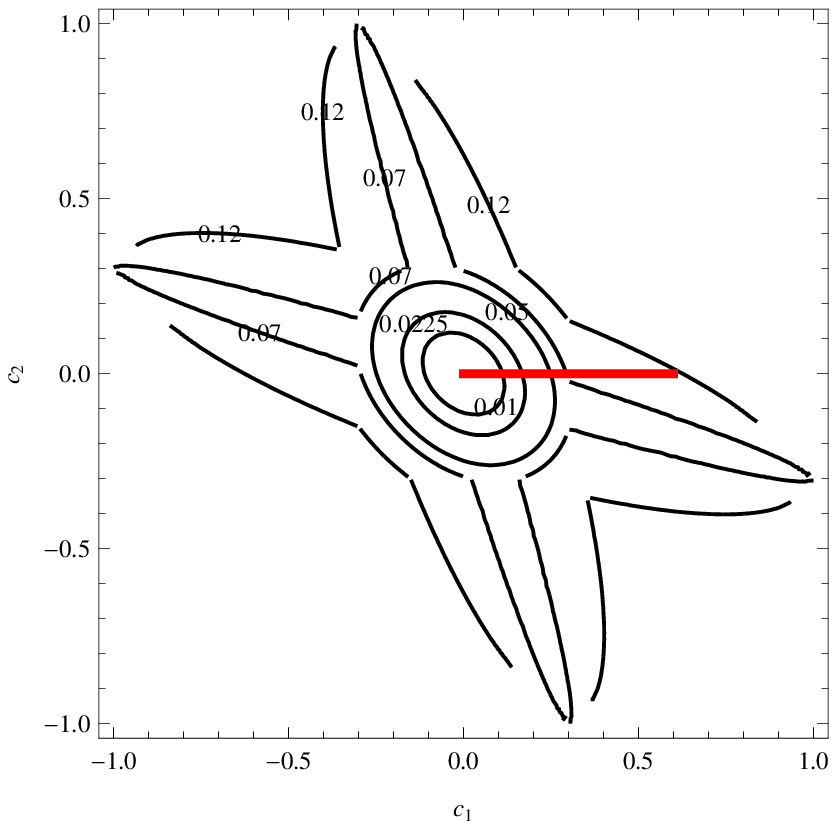}}
\subfigure[]{
\label{c}
\includegraphics[width=0.3\textwidth ]{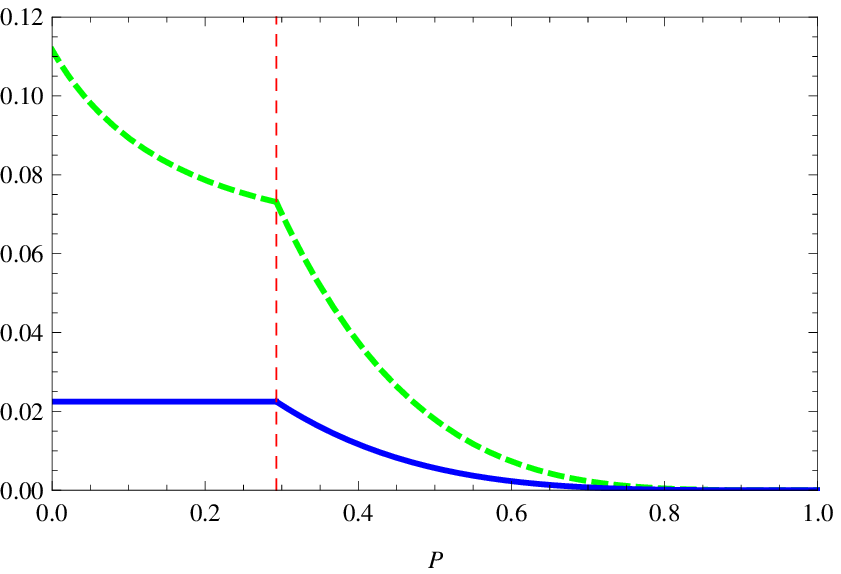}}
\end{center}
\caption{(Color online) Projections of level surfaces of geometric discord $\mathcal{D}_G$ (a) and original quantum discord
$\mathcal{D}$ (b) onto the plane $c_3=0.3$ for Bell-diagonal state with the initial conditions $c_1(0)=0.6$, $c_2(0)=0$, $c_3(0)=0.3$.
The (red) thick straight lines $\{c_1(p)=(1-p)^2c_1(0),c_2(p)=0,c_3(p)=c_3\}$ denotes the trajectories tracing out by the
Bell-diagonal state under phase flip channel.
$\mathcal{D}_G$ and $\mathcal{D}$ as a function of $p$ are also plotted (c): the (blue) solid line represents $\mathcal{D}_G$,
the (green) dashed line $\mathcal{D}$, and the (red) dashed line shows that the sudden change in decay rates of $\mathcal{D}_G$
and that of $\mathcal{D}$ occur simultaneously for Bell-diagonal states \cite{sudden change}.
}\label{projection}
\end{figure}

For clarity, we have projected the level surfaces of geometric discord $\mathcal{D}_G$ Fig. \ref{projection}(a)
and original quantum discord $\mathcal{D}$ Fig. \ref{projection}(b) onto the plane $c_3'=c_3$ for the decohering
Bell-diagonal state $\{c_1(p)=(1-p)^2c_1(0),c_2(p)=0,c_3(p)=c_3\}$ (the red straight trajectory). In Fig. \ref{projection}(a),
the trajectory coincides with one straight contour line ($\mathcal{D}_G=0.3^2/4=0.0225$) until it encounters
the sudden change point $c_1(p)=c_3$, which indicates the presence of a "decoherence-free" evolution for $\mathcal{D}_G$.
However, as depicted in the Fig. \ref{projection}(b), the same trajectory always "drills through" the contour lines, which
implies the quantum discord $\mathcal{D}$ is affected by the decoherence environment from beginning to end. Besides,
it should be noted that $\mathcal{D}_G$ and $\mathcal{D}$ share the same sudden change point for Bell-diagonal states
(see Fig. \ref{projection}(c)).

\subsection{Comparison between $\mathcal{D}_G$ and $\mathcal{D}$ for Bell-diagonal states}

In this subsection, we discuss a brief but interesting application of level surfaces of $\mathcal{D}_G$ and $\mathcal{D}$.
It is most recently numerically observed that the following hierarchical relationship holds for arbitrary
two-qubit states \cite{general state}:
\begin{eqnarray}
\label{comparison}
2\mathcal{D}_G\geq\mathcal{D}^2.
\end{eqnarray}

\begin{figure}[htbp]
\begin{center}
\includegraphics[width=.40\textwidth]{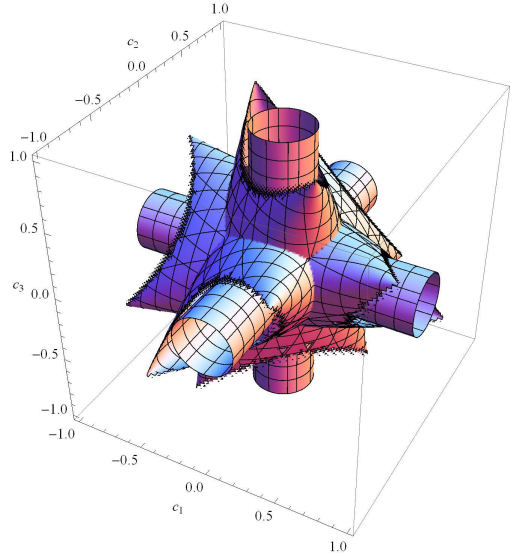} {}
\end{center}
\caption{(Color online) Contour maps of $2\mathcal{D}_G=\alpha^2$ and $\mathcal{D}=\alpha$ for a given $\alpha$
($\alpha=0.15$ in this figure and without considering $\mathcal{T}$). It is a direct illustration that contour map of $2\mathcal{D}_G=\alpha^2$ is completely inside contour $\mathcal{D}=\alpha$.
This visual spatial relation identifies Eq. \ref{comparison} for Bell-diagonal states.
}\label{relationship}
\end{figure}

In the following, we are still focusing on Bell-diagonal states for the sake of simplicity.
If we take $2\mathcal{D}_G$ as the normalized geometric discord (it is reasonable because $2\mathcal{D}_G$ and $\mathcal{D}$
simultaneously vanish for classical correlated states and reach the maximal value 1 for Bell states), then we plot the level
surfaces of $2\mathcal{D}_G=\alpha^2$ and $\mathcal{D}=\alpha$ for a given $\alpha\in[0,1]$ in the same picture
(Fig. \ref{relationship}). By varying $\alpha$ (actually changing $\mathcal{D}$) step by step, it is interestingly found that
the contour map of $2\mathcal{D}_G=\alpha^2$ is completely inside contour $\mathcal{D}=\alpha$, which means the points on the
surfaces of $\mathcal{D}$ are supposed to occupy a larger $2\mathcal{D}_G$ than $\alpha^2$, that is to say, $2\mathcal{D}_G$
must exceed $\mathcal{D}^2$ for a given point on the contour $\mathcal{D}=\alpha$.
Therefore, Eq. \ref{comparison} is verified for Bell-diagonal states from a geometric perspective. Note that the authors in
\cite{general state} can not give the upper boundary states for the geometric discord at a fixed quantum discord. However, once we acquire such a class of states, this pictorial tool is still useful to check the relationship between $\mathcal{D}_G$ and $\mathcal{D}$.

\section{Discussion and Conclusions}
In present work, we have advanced a geometric interpretation of the geometric discord and investigated the level
surfaces for this definition of quantum discord.
For an arbitrary Bell-diagonal states $\rho$ specified by $(c_1,c_2,c_3)$, the nearest zero-discord states should be
within $(c_1,0,0)$, $(0,c_2,0)$, and $(0,0,c_3)$ \cite{different measures}. In fact, the zero-discord states can be represented as
$\chi=(1+t_i\sigma_i\otimes\sigma_i)$, so the geometric discord can computed as
\begin{eqnarray}
||\rho-\chi||^2=Tr(\frac{(t_i-c_i)\sigma_i\otimes\sigma_i+c_j\sigma_j\otimes\sigma_j
+c_k\sigma_k\otimes\sigma_k}{4})^2
=\frac{(t_i-c_i)^2+c_j^2+c_k^2}{4}
\end{eqnarray}
It is obvious that $||\rho-\chi||^2$ attains the minimum when $t_i-c_i=0$. For instance,
if we assume $|c_1|\geq |c_2|,|c_3|$, thus $||\rho-\chi||^2=(c_2^2+c_3^2)/4$. Now the cylinder-like
structure of level surfaces of $\mathcal{D}_G$ is easy to be understood.

By employing this method, we have observed the dynamics of geometric discord under decoherence and interestingly
found if we expect the geometric discord to remain constant for a finite period under phase-flip channel,
the initial state must be separable. Moreover, this geometric understanding can be applied to verify the hierarchical relationships between
$\mathcal{D}_G$ and $\mathcal{D}$ for Bell-diagonal states. Our work shows that such a visualization approach can provide more clues to discover new physical phenomenon of quantum correlations, and intuitively it is a useful tool for similar geometric definitions of other physical quantities.
\section*{Acknowledgements}
This work was supported by the National Basic Research Program of China (Grants No. 2011CBA00200 and No. 2011CB921200),
National Natural Science Foundation of China (Grant No. 60921091), the National High Technology Research and Development
Program of China (863 Program) (Grant No. 2009AA01A349) and China Postdoctoral Science Foundation (Grant No. 20100480695).

\end{document}